\documentclass[prc,aps,twocolumn,superscriptaddress,preprintnumbers,amsmath,amssymb]{revtex4-2}
\usepackage[utf8]{inputenc}
\usepackage{graphicx}
\usepackage[breaklinks, colorlinks=true]{hyperref}
\usepackage{bm}
\usepackage{physics}

\newcommand{\LQCD}{\Lambda_{\mathrm{QCD}}}
\newcommand{\muB}{\mu_{\mathrm{B}}}
\newcommand{\be}{{\bm e}}
\newcommand{\bn}{{\bm n}}
\newcommand{\bp}{{\bm p}}
\newcommand{\bq}{{\bm q}}
\newcommand{\br}{{\bm r}}
\newcommand{\fm}{\,\text{fm}}
\newcommand{\MeV}{\,\text{MeV}}
\newcommand{\GeV}{\,\text{GeV}}

\begin{document}
\preprint{KEK-TH-2532, J-PARC-TH-0290, RIKEN-iTHEMS-Report-23}
\title{HBT signature for clustered substructures probing primordial
  inhomogeneity in hot and dense QCD matter}

\author{Kenji Fukushima}
\email{fuku@nt.phys.s.u-tokyo.ac.jp}
\affiliation{Department of Physics, The University of Tokyo,  7-3-1 Hongo, Bunkyo-ku, Tokyo 113-0033, Japan}

\author{Yoshimasa Hidaka}
\email{hidaka@post.kek.jp}
\affiliation{KEK Theory Center, Tsukuba 305-0801, Japan}
\affiliation{Graduate University for Advanced Studies (Sokendai), Tsukuba 305-0801, Japan}
\affiliation{Department of Physics, The University of Tokyo, 7-3-1 Hongo, Bunkyo-ku, Tokyo  113-0033, Japan}
\affiliation{International Center for Quantum-field Measurement Systems for Studies of the Universe and Particles (QUP), KEK, Tsukuba, 305-0801, Japan}
\affiliation{RIKEN iTHEMS, RIKEN, Wako 351-0198, Japan}

\author{Katsuya Inoue}
 \email{kxi@hiroshima-u.ac.jp}
\affiliation{Chemistry Program, Hiroshima University, Higashi-Hiroshima, Hiroshima 739-8526, Japan}
\affiliation{International Institute for Sustainability with Knotted Chiral Meta Matter (SKCM$^2$), Hiroshima University, Higashi-Hiroshima, Hiroshima 739-8526, Japan}
\affiliation{Chirality Research Center (CResCent), Hiroshima University, Higashi-Hiroshima, Hiroshima 739-8530, Japan}

\author{Kenta Shigaki}
 \email{shigaki@hiroshima-u.ac.jp}
\affiliation{Physics Program, Hiroshima University, Higashi-Hiroshima 739-8526, Japan}
\affiliation{International Institute for Sustainability with Knotted Chiral Meta Matter (SKCM$^2$), Hiroshima University, Higashi-Hiroshima, Hiroshima 739-8526, Japan}
\affiliation{Core of Research for the Energetic Universe (CORE-U), Hiroshima University, Higashi-Hiroshima, Hiroshima 739-8526, Japan}

\author{Yorito Yamaguchi}
 \email{yorito@hiroshima-u.ac.jp}
\affiliation{Physics Program, Hiroshima University, Higashi-Hiroshima 739-8526, Japan}
\affiliation{Core of Research for the Energetic Universe (CORE-U), Hiroshima University, Higashi-Hiroshima, Hiroshima 739-8526, Japan}

\begin{abstract}
  We propose a novel approach to probe primordial inhomogeneity in hot
  and dense matter which could be realized in non-central heavy-ion
  collisions.  Although the Hanbury~Brown and Twiss (HBT)
  interferometry is commonly used to infer the system size, the
  cluster size should be detected if substructures emerge in space.
  We demonstrate that a signal peak in the HBT two-particle
  correlation stands at the relative momentum corresponding to the
  spatial scale of pseudo one-dimensional modulation.  We assess
  detectability using the data prepared by an event generator (AMPT
  model) with clustering implemented in the particle distribution.
\end{abstract}
\maketitle

%%%%%%%%%%
\section{Introduction}
It is an unsettled problem in nuclear physics to explore the phases of
matter out of quarks and gluons.  The underlying microscopic theory
for nuclear dynamics has been established in the form of non-Abelian
gauge theory called quantum chromodynamics (QCD).
The boundaries of QCD phases in a plane of the temperature, $T$, and
the baryon chemical potential, $\muB$, constitute the QCD phase
diagram;  see
Refs.~\cite{Fukushima:2010bq, Fukushima:2013rx, Andronic:2017pug, Fischer:2018sdj}
for reviews.
As long as $\muB/T \lesssim 2$ is satisfied, the numerical Monte-Carlo
simulation of lattice-discretized QCD (i.e., lattice QCD) provides us
with reliable predictions from the first-principles
approach~\cite{Bazavov:2017dus}.
For $\muB/T \gtrsim 2$, however, the sign problem hinders the
Monte-Carlo algorithm and it still remains a major challenge to unveil
the QCD phase diagram in cold and dense regions.  There are a variety
of speculative scenarios including the QCD Critical Point, a family of
color-superconducting states, Quarkyonic
Matter~\cite{McLerran:2007qj}, dual chiral density
waves~\cite{Nakano:2004cd}, and inhomogeneous solitonic
states~\cite{Buballa:2014tba}.  In particular, some states among them
hint a certain shape of spatial modulation.  We stress that such
modulation/inhomogeneity is not bizarre but the idea of inhomogeneous
nuclear matter can be traced back to the old speculation for the
$p$-wave pion condensation~\cite{Migdal:1978az}.

If such exotic scenarios are confirmed in nuclear experiments, it would
excite wide interests beyond the nuclear community.  It has been known, 
however, that inhomogeneous phases in three spatial dimensions in 
the mean-field level are fragile against fluctuations~\cite{landau1969,peierls1934} 
and only one-dimensional (1D) quasi long-range order is expected~\cite{Hidaka:2015xza, Lee:2015bva}.  
It has been suggested that the roton-like dispersion
relation appears as a precursory phenomenon of quasi long-range order
at high enough density (called the moat regime) and the characteristic
dispersion leads to a possible experimental
signature~\cite{Pisarski:2021qof}.
We note that a stronger argument against inhomogeneous states was
given in the mean-field level in the recent study~\cite{Pannullo:2023one}. 
It is still an open question whether inhomogeneous states could exist in 
cold and dense nuclear/quark matter. 
Nevertheless, it is conceivable that clustered substructures may persist 
as a remnant which we refer to as the \textit{primordial inhomogeneity} 
with the help of strong magnetic field that effectively reduces the system 
to a pseudo 1D state in which the genuine inhomogeneity rather than 
the quasi long-range order can develop.

Now, a question is the experimental signature for the clustered substructures.
We will show that the Hanbury~Brown and Twiss (HBT)
interferometry~\cite{HanburyBrown:1956bqd} can resolve the
length scale in the particle distribution.  For a HBT related idea in the moat regime, see Ref.~\cite{Rennecke:2023xhc}.
The HBT effect is widely known as the quantum interference between
identical particles.  In nuclear experiments, it is utilized to infer
the source size of particle emission via the measured particle
correlation functions including the expanding effects~\cite{Goldhaber:1960sf,STAR:2009fks,PHENIX:2014pnh,ALICE:2011dyt}.  In the early days in relativistic heavy-ion collision physics, enhanced pion
interferometry radii were discussed as a possible consequence from a first-order phase transition from a quark-gluon plasma to the hadronic
phase~\cite{Pratt:1986cc,Bertsch:1988db,Rischke:1996em}.
The so-called ``HBT puzzle'', a counter-intuitive relation between the
sideward and the outward radii, with a na\"ive expectation with a
finite time duration of particle emission, has been intensively
discussed to be resolved~\cite{Pratt_2009}.
Recently, the technique is also applied to femtoscopic correlation
measurements to extract hadronic
interactions~\cite{STAR:2014dcy,ALICE:2020mfd}.
It is important to note that, strictly speaking, the length scale
inferred from the HBT correlation is not necessarily the size of the
whole system but the cluster size should be more relevant.  This is
usually taken as a caveat, but for our purpose to seek for
inhomogeneity, the cluster size is exactly what we pursue.

%%%%%%%%%%
\section{Primordial inhomogeneity} 
The inhomogeneous state is not robust in three spatial dimensions, but
the dimensional reduction would justify the 1D modulation.  The
well-known example is the superconductivity for which
the phase-space integral is effectively 1D near the Fermi surface.  In
the QCD context, the 1D nature at high baryon density has been
discussed in the large number of
colors~\cite{Deryagin:1992rw,Shuster:1999tn,Kojo:2009ha, Kojo:2011cn},
and the resulting inhomogeneous phase is called the Quarkyonic Chiral
Spirals~\cite{Kojo:2009ha, Kojo:2011cn}.

The dimensional reduction is further assisted by external parameters.
In the early stage in the
heavy-ion collision, the energy scale of the generated magnetic field,
$\sqrt{eB}$, reaches a scale greater than the typical QCD scale,
$\LQCD$ (or the pion mass $m_\pi$), as simulated in
Refs.~\cite{Skokov:2009qp,Deng:2012pc}, and transverse motion of
quarks is frozen. Finite-density QCD matter under strong $B$ develops
helical inhomogeneity~\cite{Basar:2010zd}, where the explicit breaking
of rotational symmetry due to magnetic field overrides the realization
of quasi long-range order. In general the lack of rotational symmetry
may lead to inhomogeneous states.

%--- figure ---%
\begin{figure}
    \centering 
    \includegraphics[width=0.7\columnwidth]{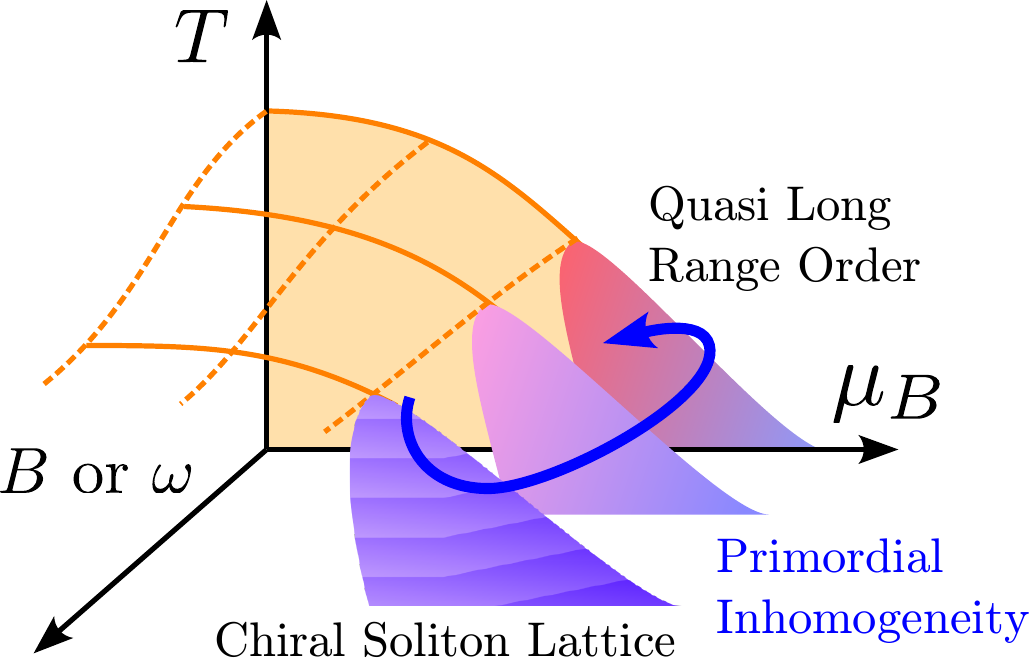}
    \caption{Schematic illustration for realization of the primordial inhomogeneity as an extension from the QCD CSL state.}
    \label{fig:schematic}
\end{figure}
% --- figure ---%

More interestingly, the low-energy effective theory of QCD under
strong $B$ can be mapped to a model for the chiral
magnet~\cite{10.1143/PTPS.159.82}.  Therefore, the QCD phase
structures can be quantitatively deduced from the phase diagram of the
chiral magnet.  In this way, an analogue of the Chiral Soliton Lattice
(CSL) is expected for
$\muB eB/(12\pi^2 f_\pi^2 m_\pi)>4/\pi$~\cite{Brauner:2016pko, Brauner:2021sci, Chen:2021vou}.
The QCD CSL state may exist in deep cores of the neutron star and in
transient matter created in the non-central (realizing strong $B$)
heavy-ion collision at intermediate energy (realizing high density).
It is pointed out that the rotation velocity $\omega$ also favors the
QCD CSL state~\cite{Huang:2017pqe}.

Let us discuss the primordial inhomogeneity.
Figure~\ref{fig:schematic} is a schematic phase diagram with an
additional axis of $B$ or $\omega$ that favors QCD CSL matter.  In
low-energy collisions the life time of the magnetic field is
significantly enhanced, and the system may transiently undergo the
CSL state.  Then, the system expands, as indicated by the arrowed
curve, toward a smaller-$B$ and dense regime where the quasi
long-range order is the true ground state. Yet, if the system
evolves sufficiently quickly, it may well be trapped in a metastable
CSL-like state, which is a mechanism to realize the primordial
inhomogeneity.

%--- figure ---%
\begin{figure}
    \centering 
    \includegraphics[width=0.7\columnwidth]{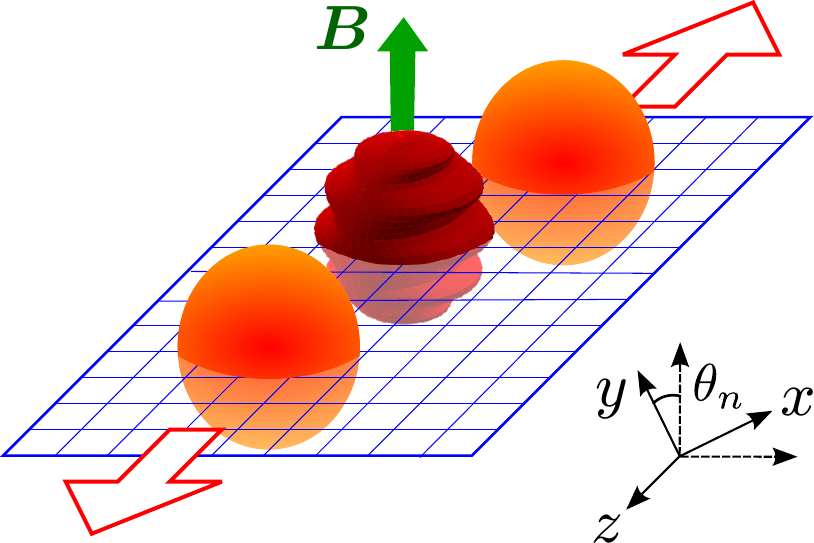}
    \caption{Collision geometry and the expected pseudo-1D modulation 
      along the magnetic direction in the non-central collision.}
    \label{fig:geometry}
\end{figure}
% --- figure ---%

The discovery of the QCD CSL state would be a intriguing challenge
that connects mathematical physics to phenomenology.
In dimensionally reduced QCD the vacuum manifold is characterized by
$U(1)_{\text{L}}\times U(1)_{\text{R}}/U(1)_{\text{V}}$,
which implies that the baryon number appears from the topological
winding from the fundamental homotopy group, $\pi_1(S^1)$, while the
baryon number arises from the $\pi_3(S^3)$ winding.  This mathematical
consideration gives feedback to phenomenology: the 1D layered sheets
of the $\pi^0$ condensate form the domain walls and the baryon number
must be localized on them.  Therefore, as illustrated in
Fig.~\ref{fig:geometry}, we can expect CSL-like pseudo-1D modulation
along the $y$-axis (which is perpendicular to the reaction plane and
parallel to $B$). Then, $\pi^0$'s and baryons could distribute in space with
layered substructures.
We note that $\pi^\pm$ are completely
suppressed in the infinitely strong-$B$ limit.  In reality, however,
the modulated $\pi^0$ is always accompanied by $\pi^\pm$ at the edges
of the domain walls~\cite{Chen:2021vou}.  So, we shall focus on the HBT
measurement for the $\pi^+$-$\pi^+$ correlation which is cleaner than
the $\pi^0$ measurement.  We need to consider the effect of the
Coulomb interaction, but the Coulomb effect is easily convoluted (or
subtracted from the experimental data) with the exact solution of the
phase shift. Therefore, assuming that the Coulomb effect is to be canceled,
we present our numerical results without any Coulomb interaction.

%%%%%%%%%%
\section{Gaussian analyses}
We define the relative momentum and the relative coordinate of two particles as
$\bq = \bp_1 - \bp_2$ and $\br=\br_1 - \br_2$.  With these variables
the two-particle correlation function can be represented as
\begin{equation}
  C_2(\bq) = \int d^3r\, S(\br) |\psi_{\mathrm{rel}}(\bq,\br)|^2
  = 1 + \langle \cos(q\cdot r) \rangle\,,
\end{equation}
where the relative wave-function is
$\psi_{\mathrm{rel}}(\bq,\br)=(e^{-i q\cdot r/2}+e^{i q\cdot r/2})/\sqrt{2}$,
so that its squared quantity is
$|\psi_{\mathrm{rel}}(\bq,\br)|^2 = 1 + \cos(q\cdot r)$, with the four vectors, $q$ and
$r$.  Using the on-shell condition, we see
that $q\cdot r$ is nothing but $-\bq\cdot \br$ in the pair rest frame.
In our convention $S(\br)$ is normalized to satisfy
$\int d^3r\,S(\br)=1$ and $\langle\cdots\rangle$ represents the
expectation value weighted by $S(\br)$.

For motivating an Ansatz for inhomogeneity in $S(\br)$, we see a
relation between $S(\br)$ and the source distribution function,
$s(\br)$.
Let us assume a simple source
function with 1D spatial modulation (which is along a unit vector
$\bn$) parametrized by
$s(\br) \propto e^{-r^2 / (2r_0^2)} [ 1 + \tilde{\alpha} \cos(2k \bn\cdot\br) ]$
apart from the normalization.  The wave-number, $k$, characterizes the
typical length scale of 1D modulation.  Then, if we make only the
back-to-back pairs (or we neglect the Lorentz boost effect which turns
out to be small), the Gaussian form is simple enough for us to
complete the integration of
$S(\br)=\int d^3 r_1\, d^3 r_2\, s(\br_1) s(\br_2) \,\delta^{(3)}(\br-\br_1+\br_2)$
in an analytical way.  The result leads us to the following
Ansatz for the modulated Gaussian:
\begin{equation}
  S(\br) = A(\alpha,k,r_0)\, e^{-r^2/(4r_0^2)}
  \bigl[ 1 + \alpha \cos(k \bn\cdot\br) \bigr] + \order{\alpha^2}\,.
\end{equation}
Here, $\alpha = 2\tilde{\alpha}\,e^{-k^2 r_0^2}$ is the amplitude of
modulation expressed in terms of parameters in $s(\br)$.
Parametrically, $\alpha$ is exponentially suppressed for
$k r_0>1$. This suppression is not a robust feature but a
consequence from a simple choice of Gaussian and cosine. Thus, we
treat $\alpha$ as a free parameter to be determined by experimental data.
The normalization constant is
$A(\alpha,k,r_0) = (4\pi r_0^2)^{-3/2} (1+\alpha\,e^{-k^2  r_0^2})^{-1}$,
with which we find
\begin{equation}
  \langle\cos(\bq\cdot\br)\rangle = 
  \frac{1 + \alpha\,e^{-k^2 r_0^2} \cosh(2 k q r_0^2)}
  {1+\alpha\,e^{-k^2 r_0^2}}\, e^{-q^2 r_0^2}
\end{equation}
for $\bn\parallel \bq$, which maximizes the modulation effect on the
HBT observable.  Now that $\bn \sim \be_y$, the optimal kinematic
condition for the modulation detection is $q_x=q_z=0$ and we construct
$C_2(\bq)$ as a function of $q_y$ only.

%--- figure ---%
\begin{figure}
    \centering 
    \includegraphics[width=0.85\columnwidth]{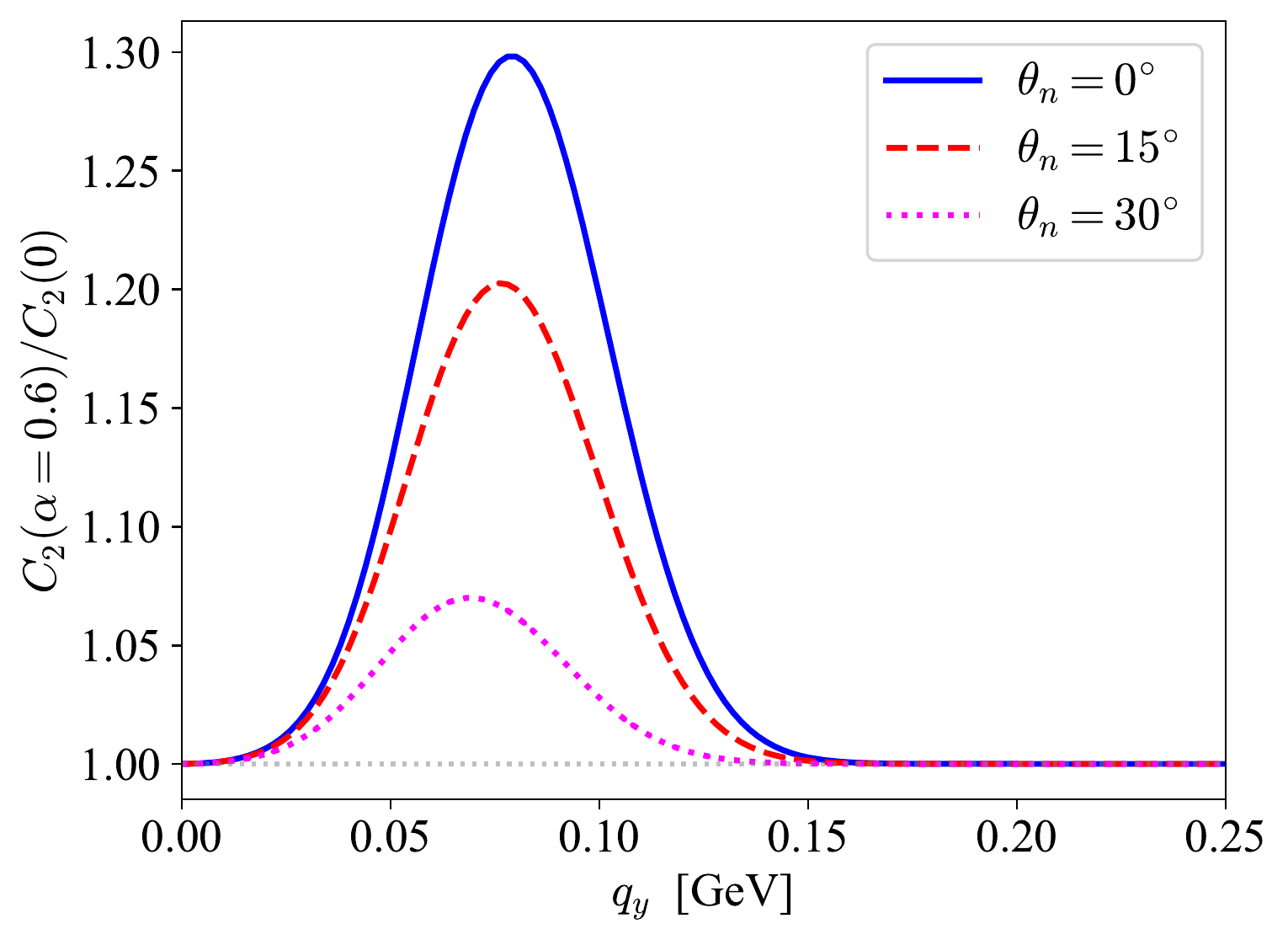}
    \caption{Normalized two-particle correlation in the simple Gaussian
      analyses.  The system size is chosen to be $r_0=6\fm$ and the
      modulation parameters are $\alpha=0.6$ and
      $k=0.4\fm^{-1}\simeq 0.08\GeV$.  The signal peak stands around
      $q_y\sim k$.}
    \label{fig:C2_analytic}
\end{figure}
%--- figure ---%

Figure~\ref{fig:C2_analytic} shows the two-particle correlation for
the parameter set, $r_0=6\fm$, $\alpha=0.6$, and $k=0.4\fm^{-1}$.
It is evident that a pronounced peak appears around $k\sim 0.08\GeV$.
We note that the typical wave number in the massless CSL is $\muB eB/(2\pi f_\pi)^2$~\cite{Brauner:2016pko} 
where $\muB$ is the baryon chemical potential.
For $eB$ comparable to $(2\pi f_\pi)^2$, the wave number $k$ should be $\sim \muB$.  
Indeed, an analogous 1D modulation, the Chiral Spirals, predicts $k\sim2\mu_{\rm B}/3$. 
If we adopt the latter relation, $k=0.4\fm^{-1}$ corresponds to $\muB\sim 120\MeV$,
i.e., $\sqrt{s_{_{NN}}}\sim 30\GeV$.  

The analytical approach is quite useful for the phenomenological
implication.  The numerical simulation is time-consuming, but we can
instantly check the parameter dependence with the obtained analytical
solution.  For example, it is practically impossible to identify the $y$-axis precisely; in other words, $\bn$ may be slightly
tilted as $\bn\cdot\be_y=\cos\theta_n\neq 1$; see the right-bottom corner in Fig.~\ref{fig:geometry}.  The sensitivity to $\theta_n$ is important in practice and, as shown in
Fig.~\ref{fig:C2_analytic}, the signal peak has strong dependence on $\theta_n$.
Also, $\alpha$ might be smaller, but our results imply that, if
$\theta_n\sim 30^\circ$ is the experimental bound, only modulations
with $\alpha\gtrsim 0.6$ are detectable by about 5\% excess in the
normalized two-particle correlation.

%%%%%%%%%%
\section{Phenomenological analyses} 
The analytical results from the Gaussian formulation are suggestive,
but we need to relax the theoretical idealization.  In analyzing
experimental data the 1D limit along the $y$-axis cannot be taken.
Thus, we must proceed to the model simulation to assess the
feasibility.  For this purpose, we adopt the AMPT
(A Multi-Phase Transport) model~\cite{Lin:2004en} to simulate the phase-space
distribution of produced particles.  More specifically, we generated
1000 events of Au-Au collisions at $\sqrt{s_{_{NN}}}=39\GeV$.  The
range of the impact parameter is $3.0\fm\le b\le 4.0\fm$ for which clustered
substructures along the $y$-axis are expected from the pseudo-1D
nature.  The modulation is introduced by hand and in this work, all the
particles are equally modulated for simplicity.  For more
systematic surveys, we should focus on particles that couple the
baryon number (such as the $\omega$ meson), but the analysis simply
goes in the same manner (with more statistics required).  The particle
distribution,
\begin{equation}
  \rho(\bp,\br,t) = \sum_n \delta(\bp-\bp_n)\,
  \delta(\br-\br_n) \delta(t-t_n)
\end{equation}
with $(\bp_n,\br_n,t_n)$ the phase-space point of $n$-th particle
emulated by AMPT{},
is shifted as
$\rho(\bp,\br-\be_y a \cos(ky),t)$
in our simple Ansatz to implement the 1D
modulation.
The modulation parameter, $k$, has the same meaning as our Gaussian
approach and let us choose $k=0.4\fm^{-1}$ again.  The amplitude $a$
is not dimensionless and we set $a=5\fm$ in this work.
Roughly speaking, the Gaussian model parameter $\alpha$ corresponds to 
$a\partial_y \rho/\rho$, where $\partial_y\rho/\rho \sim R_y^{-1}$ with $R_y$ 
the $y$-length of the system.
This parameter of $a$ is the least known part in the whole discussions
and should be related to the magnetic strength. In the future, we
should proceed to systematic investigations.  It would be an
intriguing question what $a$ is the sensitivity bound for detectability.

We mention that we mix 1000 events to make pairs.
Here, we consider the $\pi^+$-$\pi^+$ pairs and there are 416824
$\pi^+$'s from 1000 events (with the pre-selection of $p_z<1\GeV$).
Therefore, one event produces $\sim 400$ $\pi^+$'s.  If we make pairs
within each event, $\sim 8\times 10^7$ pairs are possible from 1000
events.  Since we mix 1000 events, the number of possible pairs is
$\sim 8\times 10^{10}$, which effectively corresponds to 1\,M events.

%--- figure ---%
\begin{figure}
    \centering 
    \includegraphics[width=0.85\columnwidth]{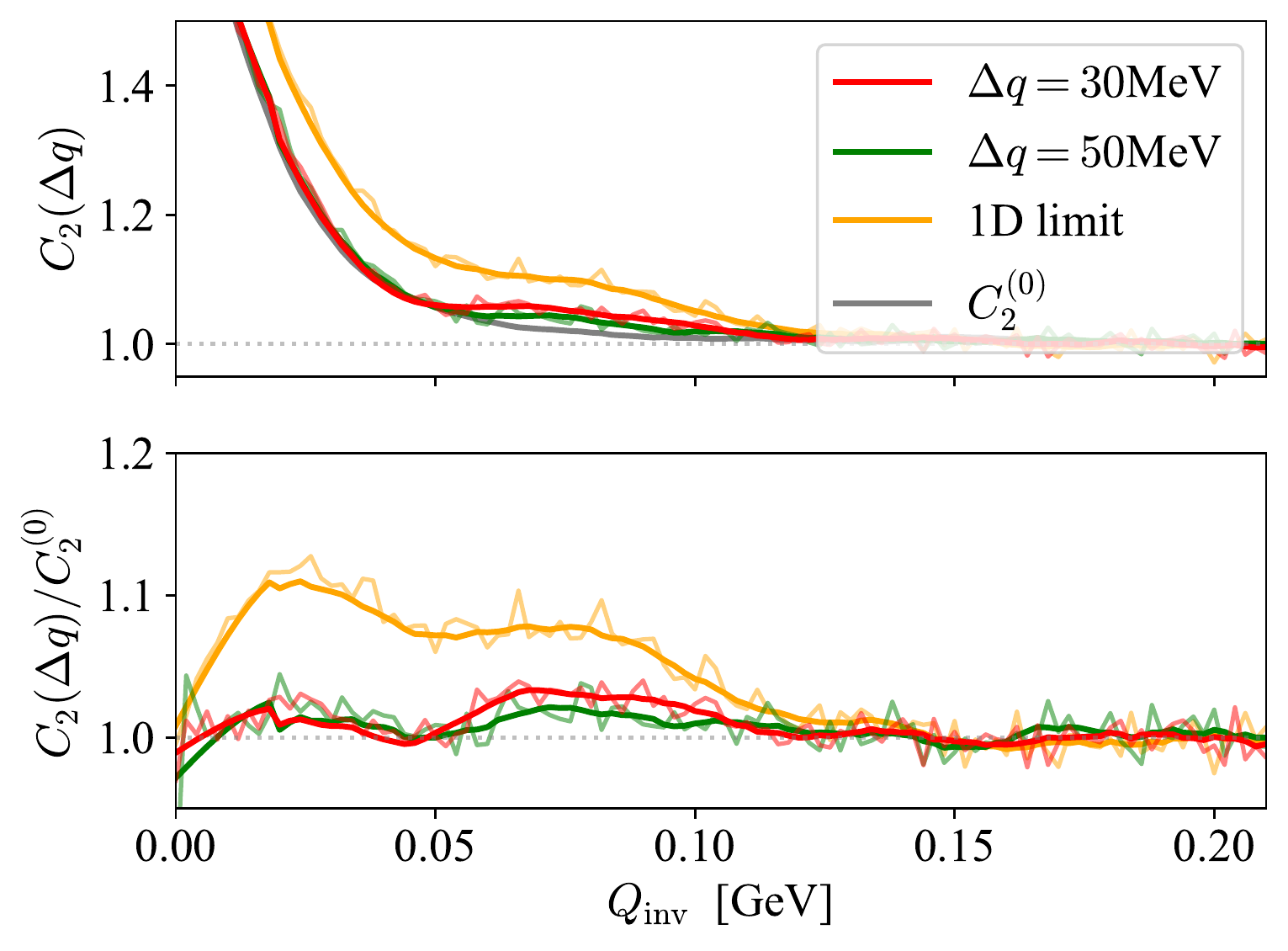}
    \caption{Two-particle correlation from the AMPT data with the
      spatial modulation.  The tilting angle is fixed as
      $\theta_n=20^\circ$.  In the upper panel $C_2^{(0)}$ is the reference and the lower panel shows the correlation normalized by $C_2^{(0)}$.}
    \label{fig:C2_AMPT}
\end{figure}
%--- figure ---%

For the evaluation of $\langle\cos(q\cdot r)\rangle$ in the transport
model calculation, $S(\br)$ is approximated into the decomposed form
of $s(\br_1) s(\br_2)$.  Then, we should make a large number of pairs,
$i$ and $j$, and make $\bq=\bp_i-\bp_j$ and $\br=\br_i-\br_j$ to take
the average of $\cos(q\cdot r)$.  We note that the boost effect to the
rest frame is included but negligibly small. The momentum filter is 
\begin{equation}
  \sqrt{q_x^2 + q_z^2} \le \Delta q\,. 
  \label{eq:filter}
\end{equation}
First, we shall consider the 1D limit of the analyses.
%It is nearly impossible to find pairs with $\bq\parallel\bn$, i.e., $q_x=q_z=0$, which corresponds to $\Delta q = 0$.
We emulate the 1D limit by computing
$\langle\cos(q_y r_y)\rangle$ instead of
$\langle\cos(q\cdot r)\rangle$ setting $q_x=q_z=0$.  Then, we see a broad bump around $0.08\GeV$ in
Fig.~\ref{fig:C2_AMPT}, for which we fix $\theta_n=20^\circ$.  For reference, the upper panel of Fig.~\ref{fig:C2_AMPT} shows $C_2$ for $\Delta q=0.3\GeV$ which is denoted by
$C_2^{(0)}$.

The lower panel of Fig.~\ref{fig:C2_AMPT} is the ratio to $C_2^{(0)}$, and 
this quantity serves as a clearer experimental signature.
The peak in raw $C_2(\Delta q)$ is washed out for large $\Delta q$, but the 
bump in the ratio remains visible by a few percent (which is experimentally 
distinguishable) even for $\Delta q=50\MeV$.
In Fig.~\ref{fig:C2_AMPT}, we present the results for both $\Delta q=30\MeV$ 
and $\Delta q=50\MeV$ to quantify the dependence on $\Delta q$.
We have numerically constructed $5\times 10^5$ pairs from 416824
$\pi^+$'s that satisfy Eq.~\eqref{eq:filter} and took the average with
the $2\MeV$ bin in terms of $Q_{\mathrm{inv}}=\sqrt{|q^2|}$.  Because $q_x$
and $q_z$ are much smaller than $q_y$ and the boost effect to the pair
rest frame is also small, the plots are hardly changed if the
horizontal axis is replaced from $Q_{\mathrm{inv}}$ to $q_y$ as in
Fig.~\ref{fig:C2_analytic}.  In Fig.~\ref{fig:C2_AMPT} the smoothed
curves over 20 data points (corresponding to the $40\MeV$ bin) are
overlaid.
In this way, we can conclude that the modulation with $a\approx 5\MeV$ is well detectable if the experimental accuracy of $\theta_n\approx 20^\circ$ is fulfilled.
It should be mentioned that we computed $C_2$ for $\theta_n=30^\circ$ and the detectability is marginal.
In this way we can make systematic assessment of detectability for a wide variation of parameters, and the present work is the first step along these lines.

%Finally, we mention that we have numerically checked the $\theta_n$ dependence.
%Figure~\ref{fig:C2_analytic} shows strong suppression for $\theta_n\neq 0$, but
%we have found that the final signal can survice.
%More specifically, we tilted the $y$-axis with $\theta=15^{\circ}$ and repeated the same calculation as in Fig.~\ref{fig:C2_AMPT}, and then we confirmed that the signal (middle curve in Fig.~\ref{fig:C2_AMPT}) is hardly affected.
%We also tested the signal for $\theta_n=30^{\circ}$, and in this case the peak disappears.
%These observations are understandable;
%as long as a peak is prominent at the level in the 1D limit as seen in Fig.~\ref{fig:C2_analytic}, the peak can persist if $\Delta q$ is sufficiently small.

%%%%%%%%%%
\section{Conclusion}
We discussed a possibility of clustered substructures in hot and dense
matter along the axis parallel to the magnetic field.
Even if the magnetic field lives short,
the pseudo one-dimensional nature in the early dynamics can induce an inhomogeneous 
density distribution and the inhomogeneity could remain afterward as a metastable state, 
which we call the primordial inhomogeneity.
We proposed a novel approach to probe the inhomogeneous state using the HBT
measurement.  Our analytical calculation in the Gaussian formalism
exhibits a pronounced peak at the relative momentum corresponding to
the wave number of spatial modulation. To assess the feasibility we
adopted the phase-space distribution of particles generated by AMPT
and computed the two-particle correlation with the spatial substructures of density distribution.  
We found that the signal excess in the correlation ratio could be suppressed by the alignment of the magnetic axis but
still persist under the appropriate momentum filter.
Our results are promising enough and the HBT correlations should deserve further systematic investigations.

%%%%%%%%%%
\begin{acknowledgments}
The authors thank
Rob~Pisarski
and
Fabian~Rennecke
for useful correspondences.
This work was supported by Japan Society for the Promotion of Science
(JSPS) KAKENHI Grant Nos.\
22H01216, 22H05118 (KF), 21H01084 (YH), 22H02053, 25220803 (KI), 18H05401 and 20H00163 (KS),
and JSPS Core-to-Core Program, A. Advanced Research Network.
\end{acknowledgments}

\bibliography{hbt}
\bibliographystyle{apsrev4-1}
\end{document}